\documentclass[a4paper]{jpconf}
\usepackage{graphicx}
\usepackage[english]{babel}
\usepackage[latin1]{inputenc}
\usepackage[intlimits]{amsmath}
\usepackage{amssymb}
\usepackage{bbm} 
\usepackage{textpos}

\newcommand{\MeV}{\,\text{MeV}}
\newcommand{\GeV}{\,\text{GeV}}
\newcommand{\s}{\,\text{s}}
\newcommand{\Mpc}{\,\text{Mpc}}
\def\stau{{\widetilde\tau}}
\def\lstau{{\widetilde\tau_1}}
\def\neut{{\chi_1^0}}
\def\hf{{X}} 
\newcommand{\OmTP}{\Omega_{3/2}^{{\rm th}}}
\newcommand{\OmNT}{\Omega_{3/2}^{\lstau}}
\newcommand{\OmSTP}{\Omega_{\hf}^{{\rm th}}}
\newcommand{\OmSNT}{\Omega_{\hf}^{\lstau}}

\newcommand{\SU}[1]{\ensuremath{\text{SU}\!\left(#1\right)}}
\newcommand{\Z}[1]{\ensuremath{\mathbbm{Z}_{#1}}} 
\newcommand{\U}[1]{\ensuremath{\text{U}\!\left(#1\right)}}

\begin{document}

\begin{textblock}{3}(10.3,-1.0) 
\noindent 
DESY 10-162
\end{textblock}

\title{Supersymmetric leptogenesis and light hidden sectors}

\author{Christoph Weniger}

\address{Deutsches Elektronen-Synchrotron (DESY), Notkestrasse 85, 22603
Hamburg, Germany}

\ead{christoph.weniger@desy.de}

\begin{abstract}
  Thermal leptogenesis and supergravity are attractive scenarios for physics
  beyond the standard model. However, it is well known that the super-weak
  interaction of the gravitino often leads to problems with primordial
  nucleosynthesis in the standard scenario of matter parity conserving MSSM +
  three right-handed neutrinos. We will present and compare two related
  solutions to these problems: 1) The conflict between BBN and leptogenesis
  can be avoided in presence of a hidden sector with light supersymmetric
  particles which open new decay channels for the dangerous long-lived
  particles. 2) If there is a condensate in the hidden sector, such additional
  decay channels can be alternatively opened by dynamical breaking of
  matter parity in the hidden sector.
\end{abstract}

\section{Introduction}
Thermal leptogenesis \cite{Fukugita:1986hr} via the out-of-equilibrium decay
of heavy right-handed neutrinos is one of the most promising models for the
generation of the baryon asymmetry in the universe. However, it is known that
this mechanism requires a high reheating temperature above $10^9\GeV$.  In
supersymmetric extensions of the standard model, such temperatures can lead to
the so-called gravitino problem, which can be circumvented when the gravitino
itself is the LSP and hence dark matter.  Indeed, it is well known that, given
a high reheating temperature $T_R\sim\mathcal{O}(10^9$--$10^{10}\GeV)$ and
gravitino masses around $m_{3/2}\sim\mathcal{O}(10$--$100\GeV)$, the thermal
relic density of gravitinos reproduces the correct dark matter abundance.

However, the above scenario is not free of problems: The next-to-lightest
supersymmetric particle (NLSP), which is often the stau, typically decays into
the gravitino with lifetimes that are of the order of minutes or days. This is
generically in conflict with the successful predictions of standard primordial
nucleosynthesis (BBN), since the NLSP decay can destroy some of the primordial
elements or lead to catalytic over-production of $^6$Li and
$^9$Be~\cite{Jedamzik:2006xz, Pospelov:2006sc}.

Different solutions to the above problem were proposed. In this note we
shortly review two possible solutions that are related to extensions of the
MSSM with light hidden sectors: 1) We will assume the existence of a hidden
sector fermion, $\hf$, lighter than the lightest observable supersymmetric
particle (LOSP). Then new decay channels are possible for the LOSP, for
instance, when the LOSP is the lightest stau or the lightest neutralino,
$\lstau \to \tau \hf$ and $\neut \to (Z^0, \gamma, h^0, f\bar f) \hf$. If
these decays are fast enough, the density of LOSPs at the time of
nucleosynthesis can be significantly reduced and thus the successful
predictions of the standard BBN scenario will not be jeopardized (for details
see Ref.~\cite{lh}). 2) It is well known in the literature that in cases where
matter parity (or $R$-parity) is weakly violated, the NLSP can decay into
standard model particles before the onset of BBN~\cite{Buchmuller:2007ui},
while being compatible with all cosmological constraints. We will present a
model which can give rise to the required small matter-parity violation (for
details see Ref.~\cite{rpv}) and discuss cosmological constraints.

\medskip

\section{Decay into hidden sector particles}
\begin{figure}[t]
  \begin{center}
    \includegraphics{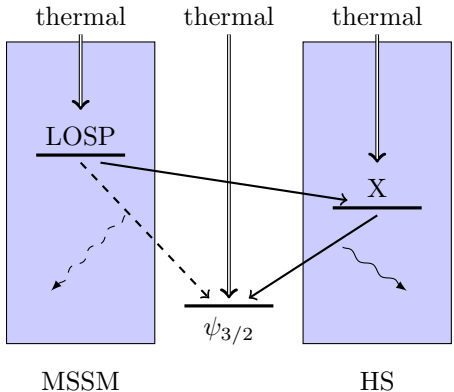}
    \hspace{0.7cm}
    \begin{minipage}[b]{9cm}
      \caption{The LOSP, the gravitino and the hidden fermion can be produced
      thermally or non-thermally through the decays of heavier particles. If
      the reheating temperature is high, $T_R\gtrsim 10^9\GeV$, the LOSP decay
      into the gravitino occurs during or after the time of primordial
      nucleosynthesis, altering the predictions of the standard BBN scenario,
      yielding abundances in conflict with observations. However, if the LOSP
      coupling to the hidden fermion is large enough, this decay can occur
      before the onset of the nucleosynthesis reactions, thus avoiding
      altogether any possible effect of the LOSP on nucleosynthesis. See
      Ref.~\cite{lh}.}
      \label{fig:sketch1}
    \end{minipage}
  \end{center}
\end{figure}

We consider a scenario where the MSSM particle content is extended with a
light standard model (SM) singlet superfield (chiral or vector). We will
further assume that the fermionic component, the \textit{hidden fermion} $X$,
of this superfield couples to the LOSP and its standard model counterpart via
a tiny Yukawa coupling, and we concentrate on the case where the LOSP is a
stau.  The interaction Lagrangian between the hidden fermion and the lightest
stau, $\lstau$, is given by the renormalizable term $-{\cal L}=\lambda_\lstau
\bar \hf\,\tau\,\lstau +{\rm h.c.}$ We will assume below that the gravitino is
the LSP and the hidden fermion $X$ the NLSP (see Ref.~\cite{lh} for other
scenarios).  The stau can then decay either via $\lstau\rightarrow
\psi_{3/2}\,\tau$ or via $\lstau\rightarrow \hf\,\tau$ with decay rates given
by $\Gamma_{\lstau \rightarrow \psi_{3/2} \tau} \sim m_{\lstau}^5 m_{3/2}^{-2}
m_P^{-2}$ and $\Gamma_{\lstau \rightarrow \hf \tau} \sim |\lambda_{\lstau}|^2
m_{\lstau}$, respectively. If the coupling $\lambda_\lstau$ is large enough,
the stau will decay before BBN, thus preventing the catalytic production of
$^6$Li. The mechanism is sketched in Fig.~\ref{fig:sketch1}.

\subsection{Cosmologically stable hidden fermions}

If the hidden fermion is stable, gravitinos as well as hidden fermions
contribute to dark matter, each of them having a thermal component and a
non-thermal component:
\begin{equation}
  \Omega_\text{dm}  = \OmTP  + \OmNT  + \OmSTP  + \OmSNT  \,.
\end{equation}
Here, $\OmNT =(m_{3/2}/m_{\lstau}) \mbox{BR}(\lstau\rightarrow\psi_{3/2}\tau )
\, \Omega_{\lstau}^{\rm th}$ and $\OmSNT =(m_{\hf}/m_{\lstau})
\mbox{BR}(\lstau\rightarrow \hf\tau ) \, \Omega_{\lstau}^{\rm th}\,,$ are the
non-thermal contributions to the gravitino and hidden fermion relic density,
respectively, where $\Omega_{\lstau}^{\rm th}$ and $\OmTP$ are the stau and
the gravitino thermal abundance. If the hidden fermion couples to the
observable sector through a renormalizable coupling, the thermal production
proceeds dominantly via the decay of thermally produced staus. The
corresponding hidden fermion relic abundance is given by $\OmSTP h^2 \sim
10^{23} |\lambda_{\lstau}|^2 (m_\hf/m_{\lstau})( 1-m_\hf^2/m_{\lstau}^2)^2$.

In order to sufficiently reduce the number density of staus at the time of BBN
it is necessary that $\text{BR}(\lstau\to\tau \hf)\simeq1$, and therefore,
$\Omega_{3/2}^{\lstau}\simeq0$. Requiring that the total dark matter density
does not exceed the measured value by WMAP implies then
$\Omega^\text{th}_{3/2} + \Omega_\hf^\text{th}
+\frac{m_X}{m_{\lstau}}\Omega_{\lstau}^\text{th} \lesssim 0.11 h^{-2}$.  In
the regime where the production of $\hf$ is sizeable, this gives a strong
upper bound of the order of $10^{-12}$ on the coupling $\lambda_{\lstau}$.

\begin{figure}[t]
  \begin{center}
    \includegraphics[width=0.5\linewidth]{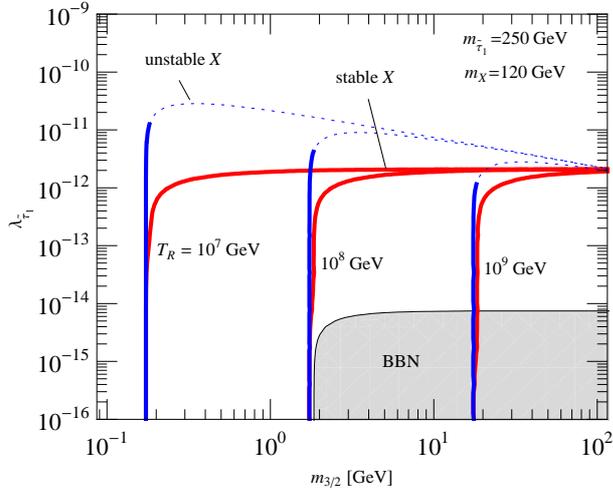}
    \hspace{0.4cm}
    \begin{minipage}[b]{7cm}
      \caption{Summary of constraints. The masses of $\hf$ and $\lstau$ are
      fixed as indicated. The \textit{red} and \textit{blue lines} show the
      values of $m_{3/2}$ and $\lambda_{\lstau}$ that yield the correct total
      relic abundance for different reheating temperatures
      $T_R=10^7$--$10^9\GeV$, assuming that $X$ is stable or unstable,
      respectively (the gluino mass has been set to $m_{\tilde g}=800\GeV$).
      Furthermore, the \textit{dashed} part of the blue lines is excluded by
      constraints on mixed warm/cold dark matter as discussed in the text, and
      the gray region is excluded by $^6$Li overproduction during BBN. See
      Ref.~\cite{lh}.}
      \label{fig:summaryI}
    \end{minipage}
  \end{center}
\end{figure}

The bounds are illustrated in Fig.~\ref{fig:summaryI} (red lines). The lines
show, for different reheating temperatures and different masses of the hidden
fermion, the value of the coupling $\lambda_\lstau$ as a function of the
gravitino mass from the requirement that the {\it total} dark matter density
is equal to the value inferred by the WMAP collaboration. For large enough
couplings $\lambda_\lstau$, the BBN bounds (gray region) can be avoided.

\subsection{Unstable hidden fermions}
If kinematically allowed, the hidden fermion $\hf$ decays into gravitinos and
hidden sector particles (which are not dangerous for BBN) well before
matter-radiation equality. In this case, dark matter consists of thermally
produced gravitinos, as well as non-thermally produced gravitinos coming from
the late decay of hidden fermions $X$ and staus $\lstau$. The dark matter
abundance is then given by
\begin{align}
  \Omega_\text{dm}=\Omega_{3/2}^\text{th} +   \underbrace{
  \Omega_{3/2}^{\lstau} +
  \frac{m_{3/2}}{m_\hf}\left(\Omega_\hf^\text{th} +
  \Omega_X^{\lstau}\right)}_{=\Omega_\text{WDM}}\;,
  \label{eqn:abu2}
\end{align}
where we assumed for simplicity that the hidden-sector particles produced in
the decay of $X$ are massless. The component coming from the late decay of
$\hf$, as well as the small fraction of gravitinos produced directly in
$\lstau$ decays, will typically act as warm dark matter (WDM), with
free-streaming lengths $\lambda_\text{FS}\gtrsim5\Mpc$. Bounds on the fraction
$f$ of the dark matter density that is allowed to be warm with a
free-streaming length above $0.5\Mpc$ were discussed in
Ref.~\cite{Boyarsky:2008xj} in the context of sterile neutrinos. There, using
Lyman-$\alpha$ data and WMAP5 results, $2\sigma$-bounds around $f\lesssim0.1$
were found for a warm component with free-streaming lengths around
$\mathcal{O}(10\Mpc)$, corresponding to $\mathcal{O}(1\ \text{km}/\s)$ thermal
velocities. Allowing a fraction $f$ of dark matter to be warm, and provided
that $\Omega_{3/2}^{\lstau}\simeq0$, implies then the upper bound
$(m_{3/2}/m_X)(\Omega_\hf^\text{th} + \Omega_\hf^{\lstau}) \lesssim f\ 0.11\
h^{-2}$.  

The allowed parameter space is shown in Fig.~\ref{fig:summaryI} (blue lines).
The dashed part of the lines is excluded by constraints on mixed cold/warm
dark matter. Compared to the scenario where the hidden fermion is stable,
larger values of the coupling $\lambda_\lstau$ are allowed.

\subsection{Experimental signatures}
In the above scenario the coupling constant of the LOSP to the hidden fermion
$X$ is typically smaller than $10^{-12}$, therefore the LOSP decay length is
much larger than the size of typical collider detectors. If the LOSP is the
lightest stau, it propagates through the detector leaving a heavily ionizing
charged track. Furthermore, if the stau velocity is small enough, it could get
trapped in the detector and decay eventually, producing a tau moving in a
non-radial direction. This signature, albeit very spectacular, is not specific
of this scenario but also arises in scenarios with stau NLSP and gravitino or
axino LSP.

However, a hint towards our proposed scenario arises from the measurement of
the coupling stau-tau-hidden fermion. More concretely, at colliders it will be
possible a determination of the stau mass and the stau lifetime, from which
the coupling $\lambda_\stau$ could be inferred. If the coupling lies in the
range preferred by cosmology, this scenario would gain strength. Further more
model dependent predictions can be found in Ref.~\cite{lh}.

\section{Dynamical matter-parity violation}
As mentioned above, a small violation of matter parity can help to reconcile
leptogenesis with gravitino dark matter. If we consider the matter parity to
be the $\Z2$ subgroup of the anomaly-free $\U1_{B-L}$ gauge symmetry, a very
natural explanation of the required smallness of the breaking is to relate it
to the condensation scale $\Lambda$ of an asymptotically free hidden sector
gauge group factor. This scale can be much lower than the $\U 1_{B-L}$
breaking scale $M_S$, and if the hidden sector is charged under $B-L$, the
condensate induces bilinear matter-parity breaking of order $\Lambda^2/M_S$.

\subsection{Model}
We consider an extension of the supersymmetric standard model with gauge group
\begin{align}
  \label{eq:G}
  G=\SU{3}_c\times \SU 2_L \times \U 1_Y \times \U 1_{B-L} \times \SU 2_{\rm
  hid} \ .
\end{align}
We assume that light neutrino masses are generated in the standard way by the
see-saw mechanism after the breaking of $\U 1_{B-L}$ at a high scale $M_S$.
Large Majorana masses for the right-handed neutrinos $N^c$ are generated by a
singlet field $S$ in our model,
\begin{align}
  \label{eq:Wss}
  W_{\text{see-saw}} =&h^{(n)}_{ij} L_i N^c_j H_u+ \frac{1}{2}\lambda^S_{i} S
  N^c_i N^c_i \ ,
\end{align}
where we have taken the basis where the Majorana mass matrix for $N^c$ is
diagonal.  We assume $\langle S \rangle = M_S$. Then, the $N_i^c$ have
Majorana masses, $M_i=\lambda^S_i\times M_S$ for $i=1-3$, respectively, and we
define $M_1<M_2<M_3$ (we will assume that $M_3=M_S$ below). For successful
thermal leptogenesis, we require $M_1> 10^9$ GeV. Since $\U 1_{B-L}$ is gauged
initially, the corresponding Nambu--Goldstone (NG) mode will be absorbed by
the gauge field, which then decouples from the low-energy effective theory.
The field $S$ necessarily has $B-L$ charge $-2$, hence a global discrete $\Z
2$ subgroup remains unbroken during this process, the  `matter parity'.  

Furthermore, the hidden sector is assumed to contain two doublet quarks
$\mathcal Q^\alpha_1, \mathcal Q^\alpha_2$ with $B-L$ charge $+1/2$, and two
doublet quarks $\mathcal Q^\alpha_3,\mathcal Q^\alpha_4$ with charge
$-1/2$.\footnote{We also require in our model the existence of five neutral
singlets $Z_{13}, Z_{14}, Z_{23},Z_{24}$ and $X$, see Ref.~\cite{rpv}.} Here
and in the following $\alpha,\beta=1,2$ are $\SU 2_{\rm hid}$ indices. The
low-energy degrees of freedom are the antisymmetric combinations $V_{ij}= -
V_{ji} = \Lambda^{-1} \mathcal Q_i^\alpha  \mathcal Q_{j \alpha}$ with
convention $ \mathcal Q_{i \alpha} = \epsilon_{\alpha \beta} \mathcal
Q_i^{\beta}$, where $\epsilon_{\alpha \beta}$ is totally antisymmetric and
$\epsilon_{12}=1$.  Condensation gives rise to non-vanishing vacuum
expectation values for the two charged effective mesons, $\langle V_{12}
\rangle = \langle V_{34} \rangle = \Lambda$ (for details see~\cite{rpv}).
Since $V_{12}$ and $V_{34}$ have $B-L$ charge $+1$ and $-1$, respectively, we
conclude that matter parity is broken dynamically at the scale $\Lambda$ in
the hidden sector, and we will take $\Lambda\ll M_S$.

The only unsuppressed and renormalizable interaction allowed by the gauge
symmetries, connecting hidden and visible sector, is given by the term
$W\supset - f_i\mathcal Q^{\alpha}_3  \mathcal Q_{4 \alpha}N^c_i$. Here, $f_i$
with $i=1,2,3$ are free parameters, and we will assume a simple situation
where $f_3\leq1$ and $f_1=f_2=0$, to show the presence of a consistent
parameter region in the model.  After SU(2)$_\text{hid}$ condensation, this
becomes a linear term for the right-handed neutrino multiplets, which together
with the mass term in Eq.~(\ref{eq:Wss}) implies a non-vanishing vacuum
expectation value for the corresponding sneutrinos,
\begin{align}
  \label{eq:Ncvev}
  \langle N^c_i \rangle = \frac{f_i}{\lambda^S_i} \frac{\Lambda^2}{M_S} \ .
\end{align}
Hence, the matter-parity breaking is mostly bilinear in our model, originating
from the Yukawa couplings in \eqref{eq:Wss}. Its scale is related to the
condensation scale of the hidden sector gauge group.  

Since the interactions between hidden and visible sector are suppressed by
$M_S$, an approximate global $\U 1_{B-L}$ symmetry in the hidden sector
remains after $S$ acquires a vev.  This $\U 1_{B-L}$ symmetry is broken by
$\mathcal Q^{\alpha} \mathcal Q_{\alpha}$ condensation, producing a nearly
massless pseudo NG multiplet in the hidden sector. After supersymmetry
breaking, soft mass terms raise the masses of the modes in the pseudo NG
multiplet. If we assume pure gravity mediation to the hidden sector, this
yields a mass for the pseudo NG boson given by $m_{a}\sim
f_3\Lambda(m_{3/2}/M_3)^{1/2} \sim 100 \MeV f_3 ( m_{3/2}/100\GeV )^{1/2}
(\Lambda/10^6 \GeV) ( M_3/10^{16} \GeV)^{-1/2}$, whereas the tree-level masses
of the fermion partner $\psi$ and the radial scalar pseudo NG component $\rho$
are given by $m_\psi \approx  m_{3/2}$ and $m_\rho \approx 4 \ m_{3/2}$. Below
we will always assume that $m_\psi>m_{3/2}$, since we are interested in
gravitino dark matter.

\begin{figure}[t]
  \begin{center}
    \includegraphics[width=0.48\linewidth]{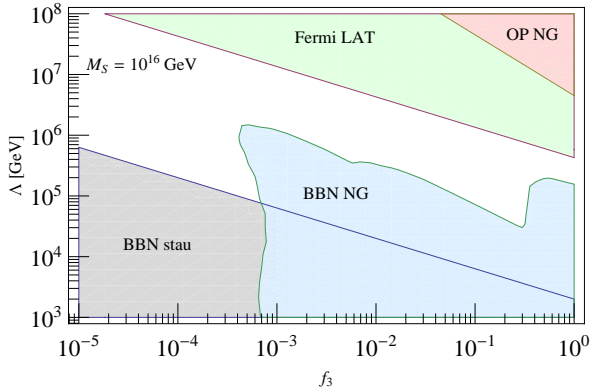}
    \hspace{0.3cm}
    \begin{minipage}[b]{7.7cm}
      \caption{Summary of constraints on condensation scale $\Lambda$.  We
      assumed $m_{3/2}=120\GeV$, $h_{ij}^{(n)}\sim10^{-1}$, $M_S=10^{16}\GeV$,
      and $T_R=10^{9.8}\GeV$.  We show the bounds coming from BBN catalysis by
      late decaying staus (BBN stau, $m_{\tilde\tau}\simeq150$--$200$ GeV),
      from the late decay of the fermionic partner of the pseudo NG boson of
      matter-parity breaking (BBN NG), and from the non-observation of
      gamma-ray lines from gravitino decay by Fermi LAT.  In the red region
      the pseudo NG boson relic density exceeds the observational limit (OP
      NG). See Ref.~\cite{rpv}.} 
      \label{fig:Summary}
    \end{minipage}
  \end{center}
\end{figure}

\subsection{\label{sec:pheno} Phenomenology}
A number of astrophysical observations constrain the lifetime of the stau and
the gravitino, which translates into bounds on the matter-parity breaking
parameters. Firstly, the $\lstau$ has to decay before BBN, with a lifetime
shorter than $2\times 10^3\s$, to avoid catalytic overproduction of
$^6$Li. This implies a lower limit on the
matter-parity breaking. Furthermore, the gravitino decay into gamma-ray lines
is limited by observations of the Fermi LAT satellite to lifetimes
$\Gamma^{-1}_{\psi_{3/2}\to\gamma\nu}\gtrsim 10^{29}\s$, yielding an upper
limit if $m_{3/2}\sim 100\GeV$. The corresponding limits on the condensation
scale $\Lambda$ are summarized in Fig.~\ref{fig:Summary} for a reference
scenario. 

Even when conservatively assuming that reheating only affects the MSSM sector,
hidden sector particles are produced in the early universe by scattering of
MSSM particles. The hidden sector contains 10 chiral multiplets with a mass at
the condensation scale $\Lambda$, as well as one chiral multiplet that remains
light and is the pseudo NG mode of the breaking of the accidental global $B-L$
symmetry in the hidden sector.  Depending on the details of the symmetries of
the hidden sector, some of the heavy hidden sector states can be stable, which
could lead to overclosure of the universe. However, depending on the details
of the model, all these potentially stable heavy hidden sector states can
either efficiently annihilate into the pseudo NG modes or decay into SM
particles.  Then, one only has to consider the effects of the NG modes (see
Ref.~\cite{rpv} for details).

The most problematic hidden sector particle turns out to be the fermionic
pseudo NG boson partner $\psi$, which mixes with the neutrinos $\nu_i$ and
inherits their SM interactions. If $m_\psi>M_Z$, the fast two-body decays
$\psi\to Z^0\nu_i$ and $\psi\to W^\pm\ell^\mp$ are kinematically allowed, but
BBN bounds are still relevant for small mixing angles and masses $m_\psi$.
The bosonic pseudo NG boson partner $\rho$ can perform two-body decays like
$\rho\to\tilde h_u \nu$, if kinematically allowed, in which case the lifetime
of $\rho$ is much smaller than the lifetime of $\psi$, since its decay is not
further suppressed by gauge couplings. The pseudo NG boson $a$ is stable on
cosmological time-scales. 

In addition to the bounds mentioned above, one can find in
Fig.~\ref{fig:Summary} bounds from BBN that come from the late decay of the
fermionic and bosonic pseudo NG boson partner. We also show the region where
the relic density of the pseudo NG boson would exceed the observational limit.
As apparent from Fig.~\ref{fig:Summary}, all constraints can be satisfied for
condensation scales $\Lambda\sim 10^5$--$10^8\GeV$.

\section{Conclusions}
Cosmological scenarios where the observed matter-antimatter asymmetry is
generated by the supersymmetric thermal leptogenesis mechanism generically
fail to reproduce the observed abundances of primordial elements. We have
discussed two possible solutions to this problem.

Firstly, we have shown that the existence of a light hidden sector fermion,
which couples very weakly to the lightest observable supersymmetric particle
(LOSP, \textit{e.g.}~the lightest stau), opens new decay channels for the
LOSP. If the coupling is large enough, the LOSP will decay dominantly into
hidden sector fermions before the epoch of primordial nucleosynthesis,
avoiding all the nucleosynthesis constraints altogether. We have summarized
the different constraints on this coupling and commented on the experimental
signatures at particle colliders.

Secondly, we have discussed a possible mechanism that generates a small
matter-parity violation in the visible sector. Due to the matter-parity
breaking, the NLSP can then decay into standard model particles before
conflicting with BBN. In our scenario, visible sector and part of the hidden
sector are simultaneously charged under a gauged $\U1_{B-L}$. This $\U1_{B-L}$
is broken at a high scale $M_S$ by right-handed neutrino masses to its
matter-parity $\Z2$ subgroup, and matter parity is then subsequently broken
completely in the hidden sector by a $\SU2_\text{hid}$ quark condensate at a
scale $\Lambda$. This mechanism gives rise to bilinear matter-parity breaking
in the visible sector of the order of $\Lambda^2/M_S$. We summarized the
constraints coming from pseudo NG modes in the  hidden sector and showed
that the model can be phenomenologically viable for large enough $B-L$
breaking scales.

\section*{References}
\bibliographystyle{iopart-num}
\bibliography{draft.bib}
\end{document}